\documentclass[useAMS]{article}

\usepackage[figuresright]{rotating}
\usepackage{amsmath}
\usepackage{natbib}
\usepackage{subfig}
\newcommand\independent{\protect\mathpalette{\protect\independenT}{\perp}} \def\independenT#1#2{\mathrel{\rlap{$#1#2$}\mkern2mu{#1#2}}}
\newcommand\eps{\hat{e_i}(C_i,\hat{\theta}_{X|C})}
\newcommand\ps{{e_i}(C_i,{\theta_{X|C}})}
\newcommand\pscut{{e_i}(C_i,{\theta^*_{X|C}})}

\newcommand\pscutnoi{{e}(C, \theta^*_{X|C})}
\newcommand\mX{m_X(X_i,C_i;\theta_{X|C})}
\newcommand\mY{m_Y(Y_i, X_i, \ps; \theta_{Y|X,e})}

\title{The central role of Bayes theorem for joint estimation of causal effects and propensity scores}

\author{Corwin M. Zigler \\
Department of Biostatistics \\
Harvard School of Public Health \\
czigler@hsph.harvard.edu
	   }

\begin{document}

\maketitle

\begin{abstract}
Although propensity scores have been central to the estimation of causal effects for over 30 years, only recently has the statistical literature begun to consider in detail methods for Bayesian estimation of propensity scores and causal effects.  Underlying this recent body of literature on Bayesian propensity score estimation is an implicit discordance between the goal of the propensity score and the use of Bayes theorem.  The propensity score condenses multivariate covariate information into a scalar to allow estimation of causal effects without specifying a model for how each covariate relates to the outcome.  Avoiding specification of a detailed model for the outcome response surface is valuable for robust estimation of causal effects, but this strategy is at odds with the use of Bayes theorem, which presupposes a full probability model for the observed data.   The goal of this paper is to explicate this fundamental feature of Bayesian estimation of causal effects with propensity scores in order to provide context for the existing literature and for future work on this important topic.
\end{abstract}

\section{Introduction}
The propensity score \citep{rosenbaum_central_1983} plays a central role in a variety of methods for estimating causal effects with observational data.  While the usefulness of the propensity score for Bayesian inference was motivated over 20 years ago \citep{rubin_use_1985}, the promise of modern Bayesian analysis tools has spawned renewed interest in Bayesian estimation of causal effects with propensity scores \citep{hoshino_bayesian_2008, mccandless_bayesian_2009,  an_Bayesian_2010, mccandless_cutting_2010, kaplan_two-step_2012, zigler_model_2013, mccandless_adjustment_2012, zigler_uncertainty_2013}.  Whereas traditional methods first estimate the propensity score, then estimate causal effects conditional on the estimated propensity score, Bayesian methods have the appeal of uniting these two estimation stages into a single analysis that jointly estimates propensity scores and causal effects.  Purported benefits of a joint Bayesian analysis include propagation of propensity score uncertainty into estimation of causal effects (vs. traditional approaches that treat the estimated quantity as fixed or rely on asymptotic arguments for variance estimation) and embedding propensity scores within broader Bayesian modeling strategies such as a model for combining data from multiple sources \citep{mccandless_adjustment_2012} or a model averaging strategy to acknowledge uncertainty in propensity score model specification \citep{zigler_uncertainty_2013}. 

Despite important progress, there exists a fundamental difficulty with unifying the two-stage nature of estimating causal effects with propensity scores into a coherent use of Bayes theorem.  This difficulty and its implications for estimating causal effects have been alluded to in existing literature in various ways and motivated a variety of quasi-Bayesian approaches, but explicit consideration of this issue is lacking.  The goal of this paper is not to propose new methodology, but rather to explicate the fundamental features of Bayesian estimation of causal effects with propensity scores in order to provide context for the existing literature and for future work on this important topic.

\section{Estimation of causal effects with propensity scores}\label{psreview}
Here we briefly review estimation of causal effects with propensity scores.  Consider $n$ observational units that have been randomly sampled from a population of interest.  For each unit, we measure a continuous, categorical, ordinal, or time-to-event outcome, $Y_i$, for $i=1,2, \ldots, n$.  Let $X_i$ be a binary variable with $X_i=1$ denoting exposure to a treatment of interest and $X_i=0$ otherwise.  Generalizations to non-binary and/or time-varying treatments are available, but we forego these extensions to simplify presentation.  The goal is to estimate the causal effect of exposure to $X=1$ (vs. $X=0$) on $Y$.  This goal is typically complicated in observational settings because assignment to treatment is not randomized.  Let $C_i=C_{i1,} C_{i2}, \ldots, C_{ip}$ denote a vector of $p$ pre-treatment variables that simultaneously relate to the receipt of treatment and to the outcome of interest, that is, let $C_i$ be the vector of confounders measured for $i=1,2, \ldots, n$. 

\subsection{The rationale for using propensity scores to estimate causal effects}
The key assumption underlying estimation of causal effects with propensity scores is that of \textit{strongly ignorable treatment assignment} stating that conditional on $C_i$, assignment to $X_i$ is conditionally independent of the potential values of $Y_i$ that would be observed under all possible levels of $X_i$ \citep{rosenbaum_central_1983}.  We forego the use of potential-outcomes notation here, and simply describe the assumption of strongly ignorable treatment assignment as formalization of the assumption that there is no unmeasured confounding.

The propensity score for the $i^{th}$ unit is the conditional probability of exposure to $X_i=1$, conditional on $C_i$:  $e_i \equiv p(X_i=1 | C_i)$. In observational studies, the propensity score is unknown and must be estimated.  Denote the estimated propensity score with $\hat{e_i}$.

The rationale for estimating causal effects with propensity scores rests on the result that $e_i$ is a \textit{balancing score} so that $X_i \independent C_i |$ $e_i$; conditional on the propensity score, the treatment indicator is independent of the covariates.  Under strong ignorability, this allows average outcome comparisons among observations with the same value of $e_i$ to serve as estimates of the average causal effect at that value of $e_i$.  The balancing-score property motivates a variety of methods for estimating causal effects whereby the estimated quantities $\hat{e_i}$ are obtained and then causal effects are estimated by treating $\hat{e_i}$ as a covariate, matching or subclassifying units by $\hat{e_i}$, or weighting by a function of $\hat{e_i}$ \citep{rosenbaum_central_1983, robins_marginal_2000, lunceford_stratification_2004, stuart_matching_2010}. 

\subsection{Models for estimating propensity scores and causal effects}\label{psmods}
Throughout, we use $\theta$ to represent parameters indexing models for observed data, with different subscripts indicating that parameter vectors are distinct depending on different model specifications. Let $\mX$ denote a model for $p(X_i=1|C_i)$, indexed by a parameter $\theta_{X|C}$.  Estimates from this model for the treatment assignment mechanism (i.e., ``the propensity score model'') yield $\hat{e}_i$ . Note that specification of $\mX$ can vary in flexibility from commonly-used logistic regression to modern machine-learning specifications such as boosted regression trees \citep{mccaffrey_propensity_2004}.   Henceforth, we augment notation to explicate the link between the propensity score and $\mX$ and denote the propensity score and its estimate with $\ps$ and $\eps$, respectively.  When $\mX$ accurately reflects the treatment assignment mechanism, $\eps$ enjoys the property of a balancing score. 

We take the estimand of interest as the average causal effect, denoted with $\Delta$ and defined as:
\begin{align}
\Delta \equiv E_C[E[&Y_i|X_i=1,C_i] - E[Y_i|X_i=0,C_i]] = \notag \\
&E_{\ps}[E[Y_i|X_i=1,\ps] - E[Y_i|X_i=0,\ps]]
\end{align}
Various strategies are available for estimating $\Delta$ with propensity scores, but we consider those where a model is specified to relate $Y$ to $\ps$ (i.e., ``the outcome model'').  We discuss non-model-based strategies in Section \ref{discussion}.  Denote the outcome model with $\mY$, which could simply specify a regression model with covariate adjustment for $\ps$, or correspond to a flexible parameterization that estimates a different causal effect for different subgroups (e.g., quintiles or matched subclasses) of observations having similar values of $\ps$. 

Traditional estimation of $\Delta$ would follow from first specifying $\mX$ and estimating $\theta_{X|C}$ to obtain $\eps$ and then using $m_Y(Y_i,X_i,\eps;\theta_{Y|X,e})$ to estimate $\theta_{Y|X,e}$ and obtain an estimate of $\Delta$, that is, estimation of causal effects with the outcome model is conducted conditional on the estimated propensity score.   To focus presentation, we forego discussion of other important steps to a propensity score analysis such as checking covariate balance or using the propensity score to preprocess the sample before estimating causal effects \citep{ho_matching_2007}.

\section{Bayesian Estimation of Causal Effects with Propensity Scores}\label{bayesps}
While \cite{rosenbaum_central_1983} note that estimates of the propensity score can be construed in a Bayesian paradigm as posterior-predictive probabilities of assignment to $X=1$, they do not consider how to incorporate these posterior-predictive probabilities into estimation of causal effects. 

For Bayesian inference, a model, however flexible, must be specified for the joint probability distribution of all potentially observable quantities, here $\mathbf{(X,Y,C,\theta)}$, where boldface letters represent vectors and matrices for observed quantities for all $n$ units and $\theta$ denotes a fixed but unknown generic parameter vector. For application of Bayes theorem, this joint distribution is factored into an expression for the likelihood, $p(\mathbf{X,Y,C}|\theta)$, describing how the data are generated conditional on the true $\theta$, and a prior distribution, $p(\theta)$, describing \textit{a priori} beliefs about possible values of the true $\theta$.  Information obtained in the data updates $p(\theta)$ to obtain a posterior distribution of $p(\theta|\mathbf{X,Y,C})$ that can be used to estimate $\Delta$.  The posterior distribution of $\Delta$ can be expressed as:
\begin{equation}
p(\Delta | \mathbf{X, C, Y}) \propto \int \Delta(\mathbf{X, Y, C}, \theta) \prod_{i=1}^n p(X_i, Y_i, C_i|\theta)p(\theta)d\theta, \label{ACEpost}
\end{equation}
where  $\Delta(\mathbf{X, Y, C}, \theta)$ denotes that, conditional on $\theta$, the quantity $\Delta$ is a deterministic function of $(\mathbf{X,Y,C}, \theta)$, and $\prod p(X_i, Y_i, C_i|\theta)p(\theta)$ is proportional to the posterior distribution $p(\theta|\mathbf{X, Y, C})$.  

The key modeling steps for Bayesian analysis are specification of a prior distribution, $p(\theta)$, and a model for $p(X_i, Y_i, C_i|\theta)$ constituting the likelihood function. Throughout, we take the marginal distribution of $\mathbf{C}$ as its empirical distribution, and omit this distribution from likelihood expressions to simplify notation.  Factorization of $p(X_i, Y_i, C_i|\theta) = p(X_i|C_i, \theta_{X|C})p(Y_i|X_i,C_i,\theta_{Y|X,C})$ seems to correspond naturally to the two stages of estimating causal effects with propensity scores; $p(X_i|C_i, \theta_{X|C})$ is the propensity score model and $p(Y_i|X_i,C_i,\theta_{Y|X,C})$ is the outcome model that does not adjust for $C_i$ directly but is reparameterized to adjust for a function of $C_i$, namely, $\ps$.  This would imply the following likelihood expression for estimating the posterior distribution of $\Delta$:
\begin{equation}
 \prod_{i=1}^n p(X_i|C_i,\theta_{X|C})p(Y_i|X_i, \ps, \theta_{Y|X,e}), \label{wronglike}
\end{equation}
where $p(X_i|C_i, \theta_{X|C})$ and $p(Y_i|X_i,\ps,\theta_{Y|X,e})$ are modeled with $\mX$ and $\mY$, respectively.  Note that the parameter $\theta_{Y|X,e}$ is distinct from $\theta_{Y|X,C}$.

Bayesian estimation of causal effects with (\ref{wronglike}) was proposed in \cite{mccandless_bayesian_2009}, but shown in \cite{zigler_model_2013} to yield biased estimates of causal effects without further adjustments. That is, the posterior distribution in (\ref{ACEpost}) cannot, in general, be recovered with a likelihood specified as in (\ref{wronglike}).   

\subsection{The fundamental problem of joint Bayesian estimation of propensity scores and causal effects}\label{fundamental}
The goal of estimating the propensity score is to condense the $p$-dimensional covariate vector, $C_i$, into a scalar quantity that can be used to adjust for confounding.  Specifically, $C_i$ is condensed into $\ps$ in accordance with the treatment assignment mechanism so that $\ps$ is a balancing score that permits estimation of $\Delta$.  The primary appeal of this approach is that it \textit{specifically precludes the need to specify a model for the entire covariate-outcome response surface.}  That is, adjustment for the propensity score is not meant to reflect a model for $p(Y_i|X_i,C_i, \theta_{Y|X,C})$; it is simply meant to facilitate comparisons between $[Y|X=1]$ and $[Y|X=0]$ that are not confounded by $C$.  Accurately modeling $p(Y|X,C, \theta_{Y|X,C})$ without $\ps$ would, of course, accomplish this goal as well, but the propensity score is typically employed in settings where the researcher wishes not to attempt this modeling exercise.  Foregoing a model for the entire covariate-outcome response surface motivates the appeal of estimating causal effects with propensity scores, but this fundamental benefit has profound implications for the application of Bayes theorem to obtain the posterior distribution in (\ref{ACEpost}).    

Because the propensity score is not meant to recover $p(Y_i|X_i,C_i,\theta_{Y|X,C})$, the fundamental problem for using Bayes theorem to combine estimation of propensity scores with estimation of causal effects can be succinctly stated in the following expression:
\begin{equation}
p(\mathbf{X,Y,C}|\theta) \ne p(\mathbf{X | C}, \theta_{X|C})p(\mathbf{Y|X,e(C},\theta_{X|C}),\theta_{Y|X,e}),
\end{equation}
i.e., a model for how $\ps$ relates to $Y$, however flexible, cannot in general serve the same purpose in Bayesian analysis as a model for the entire covariate-outcome response surface.  The reason is that the posterior distribution in (\ref{ACEpost}) relies on factorizing the entire joint conditional distribution of $p(\mathbf{Y, X, C} | \theta)$. This distribution cannot generally be factorized as $\prod p(X_i|C_i, \theta_{X|C})p(Y_i|X_i, \ps, \theta_{Y|X,e})$ because use of $p(Y_i|X_i,\ps,\theta_{Y|X,e})$ is specifically designed to avoid specification of key features of this joint distribution.  Use of the likelihood in (\ref{wronglike}) does not correspond to a valid use of Bayes theorem because it does not correspond to a coherent factorization of the joint distribution of the observed data, conditional on $\theta$.  

One practical consequence of this fundamental issue was described in \cite{zigler_model_2013} as ``feedback'' of information from $p(Y_i|X_i,\ps, \theta_{Y|X,e})$ into estimation of $\theta_{X|C}$ and, in turn, $\ps$.  This phenomenon was shown to provide invalid inferences for $\theta_{X|C}$, which distorts the balancing-sore property of $\ps$ and yields incorrect estimates of $\Delta$, confirming that a likelihood such as that in (\ref{wronglike}) cannot be used to recover the posterior distribution of $\Delta$ in (\ref{ACEpost}).  However, \cite{zigler_model_2013} also show that augmenting the outcome model to be parameterized in terms of both $\ps$ and $C_i$ directly does permit estimation of $\Delta$ with (\ref{ACEpost}). This strategy implies a joint distribution for $\mathbf{(Y,X,C)}$, and also benefits from accurate characterization of $\ps$ for estimating causal effects.  \cite{zigler_uncertainty_2013} used an augmented outcome model adjusting for both $\ps$ and $C_i$ to embed estimation of causal effects with propensity scores within a Bayesian framework for variable selection and model averaging that acknowledges uncertainty in propensity score model specification.

\subsection{Quasi-Bayesian Procedures}\label{cutbayes}
There are a multitude of approaches for Bayesian estimation of causal effects with propensity scores that maintain the inherent two-stage nature of first estimating the propensity score then estimating $\Delta$, conditional on the estimated propensity score. \cite{rubin_use_1985} describes Bayesian inference that conditions entirely on $\eps$, without regard to how $\eps$ is obtained.  \cite{rosenbaum_central_1983} note that $\eps$ can be construed as posterior-predictive probabilities in the Bayesian paradigm, which could then be used to estimate causal effects.  In essence, the use of Bayesian methods in either the propensity score modeling stage or the outcome modeling stage is straightforward, but combining these two stages into a single analysis is met with the issues discussed in Section \ref{fundamental}.

Various methods described as ``quasi-Bayesian,'' ``approximately-Bayesian,''  or ``two-step Bayesian'' have been recently proposed to combine estimation of the propensity score with that of causal effects without appealing to Bayes theorem to unify these two stages with a joint likelihood \citep{hoshino_bayesian_2008, mccandless_cutting_2010, kaplan_two-step_2012, zigler_uncertainty_2013}.  These methods have been described as ``cutting the feedback'' between the propensity score and outcome stages \citep{lunn_combining_2009}, and represent a special case of so-called ``modularization'' in Bayesian inference \citep{liu_modularization_2009}.  These strategies obtain the posterior-predictive propensity score by simulating $\theta^*_{X|C}$ from the posterior predictive distribution $p(\theta_{X|C} | \mathbf{X, C})$, without conditioning on $Y$ or other quantities in the outcome model.  Since the propensity score is a deterministic function of $(C_i, X_i, \theta_{X|C})$, $\theta^*_{X|C}$ imply simulations of the posterior-predictive propensity scores, denoted with $e_i(C_i, \theta_{X|C}^*)$,
%\begin{equation}
 %p(e_i^* | X, C) \equiv \int \ps p(X|C, \theta_{X|C})p(\theta_{X|C})d\theta_{X|C} \label{pspostpred}
%\end{equation}
which are used to estimate causal effects with the quasi-posterior distribution:
%\begin{equation}
%p(\Delta | \mathbf{X, \pscut, Y}) \propto \int p(\Delta | \mathbf{X, Y, \pscut}, \theta) \prod_{i=1}^n p(Y_i|X_i, \pscut|\theta)p(\theta)d\theta. \label{cutout}
 %\end{equation}
% \begin{align}
%p(\Delta& | \mathbf{X, \pscut, Y}) \propto \int  p(\Delta | \mathbf{X, Y, \pscut}, \theta_{Y|X,e^*}) \notag  \\
 %&\times  \prod_{i=1}^n p(Y_i|X_i, \pscut,\theta_{Y|X,e^*})p(\theta_{X|C}|X_i,C_i)p(\theta_{X|C})p(\theta_{Y|X,e^*})d\theta_{X|C} d\theta_{Y|X,e^*}. \label{cutout}
% \end{align}

\begin{align} 
p(\Delta &| \mathbf{X, C, Y}) \propto \label{cutout} \\
  &\int \Delta(\mathbf{X,Y,\pscutnoi}, \mathbf{\theta_{Y|X,e^*}}) p(\mathbf{Y|X, \pscutnoi, \theta_{Y|X,e^*}}) \notag  \\ 
 \times \Big[&\int \mathbf{e(C,\theta_{X|C})} p(\mathbf{\theta_{X|C}|X,C})d\mathbf{\theta_{X|C}} \Big]\notag \\
 \times &p(\mathbf{\theta_{Y|X,e^*}})d\mathbf{\theta_{Y|X,e^*}}, \notag
\end{align}
where $\theta_{Y|X,e^*}$ is distinct from $\theta_{Y|X,e}$ in expression (\ref{wronglike}) because $\theta_{Y|X,e^*}$ indexes a model that is conditional on the posterior-predictive propensity score obtained without regard to $Y$.  

Note that the interior integral in expression (\ref{cutout}) with respect to $\mathbf{\theta_{X|C}}$ represents the posterior-predictive propensity score distribution, which is obtained independently of any model specification for $Y$ . Thus, estimating causal effects with (\ref{cutout}) maintains the two-stage nature of estimating causal effects with propensity scores and does not appeal to Bayes theorem to combine these stages, nor does it attempt to model the entire joint distribution of $\mathbf{(X,Y,C)}$.   This procedure parallels that of multiple imputation, where, say $m$ samples of $\pscut$ are obtained according to the posterior predictive distribution in the interior integral of expression (\ref{cutout}), treated as a ``distributional constant,'' and iteratively plugged in to the outcome model to obtain $m$ samples from a quasi-posterior distribution of $\Delta$ while reflecting uncertainty in the estimated propensity score.  Additionally, these methods strictly adhere to one philosophical motivation that the propensity score, as a tool meant to mimic the design of a randomized study, should be estimated separately from the analysis of causal effects \citep{rubin_for_2008}.

%\subsection{Estimating causal effects without an outcome model}\label{nomodel}

\section{Illustration with Simulated Data}\label{sim}
The consequences of the issues described in Section \ref{bayesps} can be illustrated with a simple simulation study.  Data sets are simulated with binary $X_i, Y_i$ and $p=6$ covariates $C_i = C_{i1}, C_{i2}, \ldots, C_{i6}$ for $i=1, 2, \ldots, 1000$.  For each replicated data set, $X_i$ are simulated from Bernoulli distributions with probability of success generated from the following logistic regression: $logit(P(X_i=1|C_i)) = 0.1C_{i1} + 0.2C_{i2} + 0.3C_{i3} + 0.4C_{i4} + 0.5C_{i5} + 0.6C_{i6}$.  Similarly, $Y_i$ are simulated from Bernoulli distributions with probability of success generated from another logistic regression specified as: $logit(P(Y_i=1|C_i, X_i)) = 0.6C_{i1} + 0.5e^{C_{i2}-1} + 0.4C_{i3} + 0.3e^{C_{i4}-1} + 0.2|C_{i5}| + 0.1|C_{i6}|$.  

Following the approach described in Section \ref{psmods}, the simulated data sets are analyzed with $\mX$ specified as a logistic regression with a main effect for each component of $C_i$:
\begin{align}
logit(P(X_i=1|C_i)) = &\theta_{X|C.0} + \sum_{j=1}^6 \theta_{X|C.j}C_{ij}
\end{align}
Thus, the propensity score model used for analysis correctly reflects the true treatment assignment mechanism.  For the outcome model, we present analyses with two different outcome models.  First, we use a model with a main effect for $X_i$ and dummy variables indicating membership in quintiles of $\ps$: 
\begin{align}
logit(&P(Y_i=1|X,\ps)) = \theta_{Y|X,e.0} + \theta_{Y|X,e.1}X_i + \sum_{k=2}^5 \theta_{Y|X,e.k}I(\ps \in q_k) \label{psonly}
\end{align}
where $I(\ps \in q_k)$ takes on the value 1 if $\ps$ is in the $k^{th}$ quintile bin and 0 else.  In accordance with Section \ref{fundamental}, we also analyze the simulated datasets with a logistic regression that augments (\ref{psonly}) above to adjust for both $\ps$ and $C_i$ directly: 
\begin{align}
logit(P(Y_i=1|X,\ps)) = &\theta_{Y|X,e.0} + \theta_{Y|X,e.1}X_i \notag \\
+ &\sum_{k=2}^5 \theta_{Y|X,e.k}I(\ps \in q_k) + \sum_{j=1}^6 \theta_{Y|X,e.5+j}C_{ij} \label{psandc}
\end{align}
Note that the direct covariate adjustment in (\ref{psandc}) does not specify the transformed versions of $C_i$ that were used to simulate the data, i.e., the linear adjustment term is not a correct representation of the data generating mechanism.  

\begin{figure}
\caption{Point estimates obtained from analysis of 1000 replicated simulated datasets according to the description in Section \ref{sim}.  $^*$ denotes estimates of $\Delta$ from Strategy A2.}\label{compare}
\includegraphics[width=\textwidth]{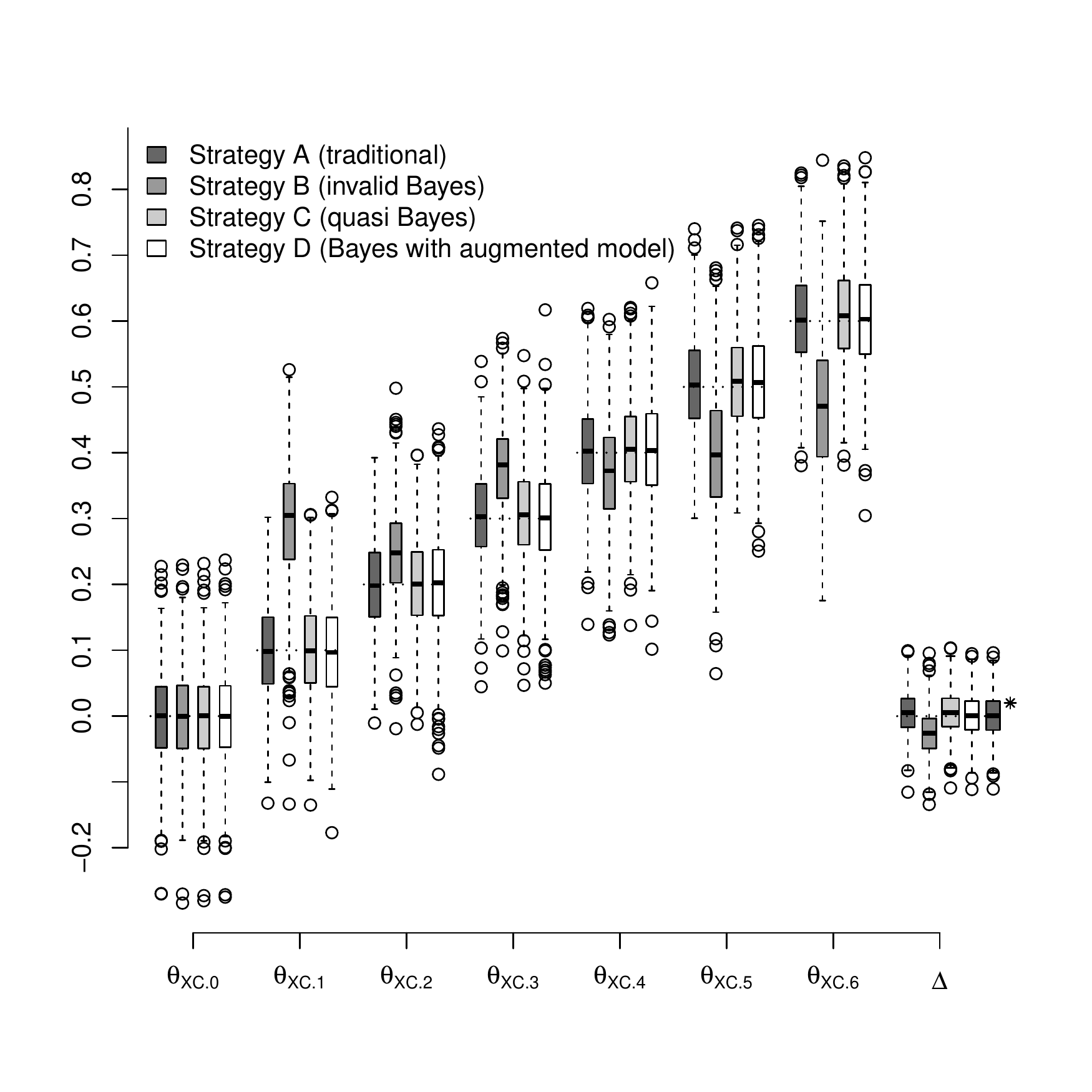}
\end{figure}

A total of 1000 data sets are simulated and each is analyzed with five strategies: Strategy A1 uses a traditional sequential approach described in Section \ref{psmods} where maximum likelihood estimates are used to estimate $\eps$ and the outcome model (\ref{psonly}) is used to estimate $\Delta$; Strategy A2 uses the same traditional sequential approach but with the outcome model specifying additional covariate adjustment as in (\ref{psandc});  Strategy B uses a Bayesian approach with the likelihood in (\ref{wronglike}) and outcome model (\ref{psonly}), which does not correspond to a valid factorization of the joint distribution of the data; Strategy C uses a quasi-Bayesain approach described in Section \ref{cutbayes} with outcome model (\ref{psonly}); and Strategy D uses the Bayesian approach described in Section \ref{fundamental} that uses outcome model (\ref{psandc}).  Results from the simulated data sets are summarized in Figure \ref{compare}, which depicts box plots of point estimates of the parameters in the propensity score model, $\theta_{X|C}$, and point estimates of $\Delta$ (maximum likelihood for the traditional approach, posterior-mean estimates for the others).  Note that the propensity score estimates, $\theta_{X|C}$ are identical for Strategies A1 and A2 described above, and are only pictured once.  

\begin{figure}
\caption{Uncertainty interval widths for estimates of $\Delta$ obtained from analysis  of 1000 replicated simulated datasets according to the description in Section \ref{sim}.}\label{compare_CI}
\includegraphics[width=.5\textwidth]{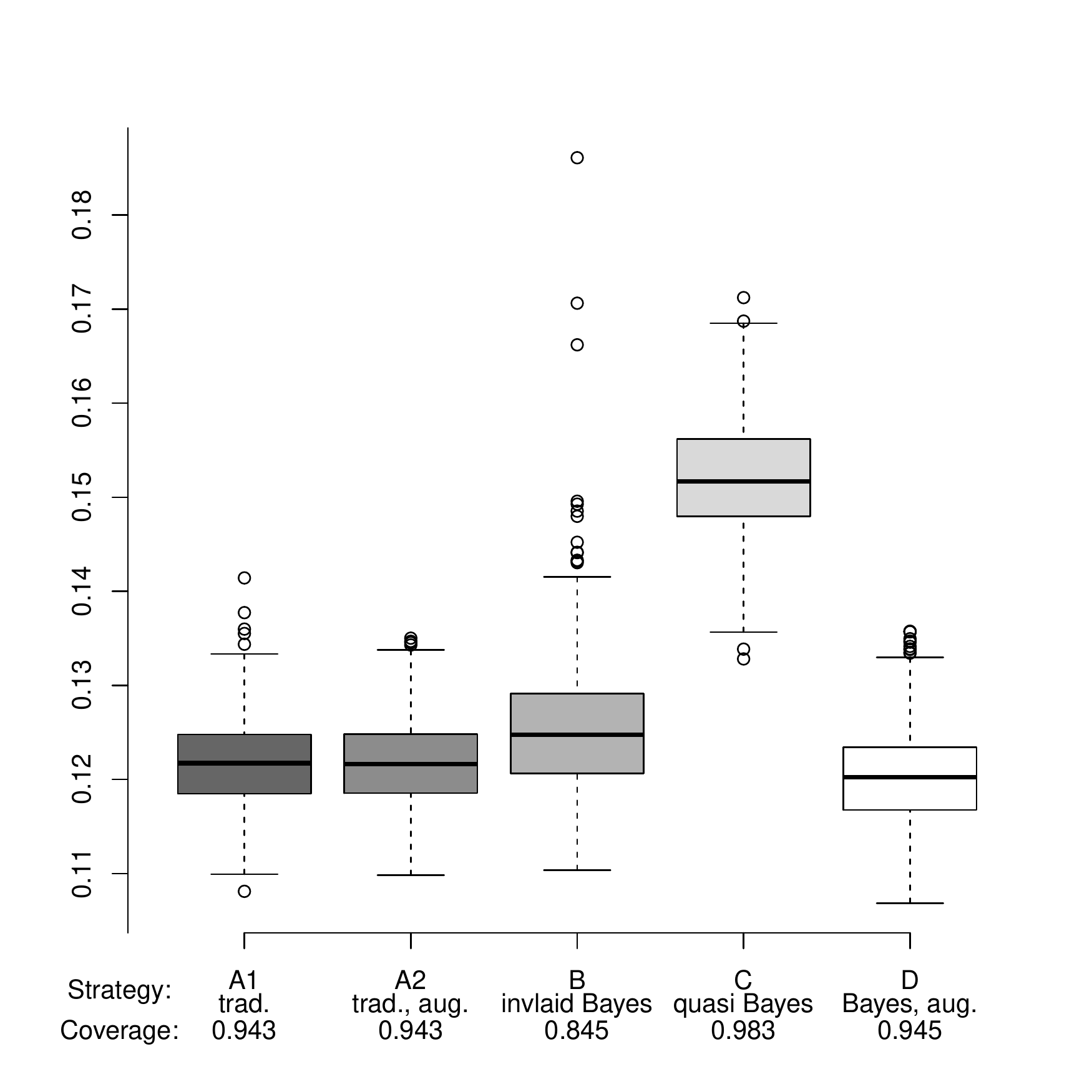}
\end{figure}

Figure \ref{compare_CI} depicts box plots of the widths of 95\% uncertainty intervals for estimates of $\Delta$, along with coverage probabilities.  For the traditional sequential strategies A1 and A2, these correspond to confidence intervals estimated from 1000 bootstrap samples for the analysis of each simulated data set, for all other strategies these correspond to 95\% posterior intervals from the analysis of each simulated data set.  

From Figure \ref{compare}, the failure of Strategy B that uses the likelihood in (\ref{wronglike}) is clear; estimates of $\theta_{X|C}$ do not correspond to the true parameters of the propensity score model, and the resulting point estimates for $\Delta$ are clearly biased.  In contrast, both the quasi-Bayesian approach (Strategy C) and the approach that augments the outcome model adjusting for $\ps$ with additional covariate adjustment (Strategy D) accurately characterize the treatment assignment mechanism and estimate $\Delta$ with performance that is comparable to the analogous traditional procedures that sequentially estimate $\eps$ and then estimates $\Delta$, conditional on $\eps$ (Strategies A1 and A2).  From Figure \ref{compare_CI} we see that Strategies A1, A2, and D exhibit coverage near the nominal rate, withe Strategy D tending to yield the narrowest uncertainty intervals.  The quasi-Bayesian Strategy C exhibits the widest intervals and coverage above the nominal rate.  The invalid Bayesian approach of Strategy B exhibits poor coverage.

\section{Conclusion}\label{discussion}
Modern tools for Bayesian analysis hold important promise for estimating causal effects with propensity scores, but there exists a fundamental discordance between the need for a fully Bayesian analysis to specify the joint distribution of all observable data (conditional on parameters) and the propensity score's goal of estimating causal effects without modeling the joint distribution of outcome and covariates.  Awareness of this discordance is necessary for understanding how Bayesian methods have been and will continue to be used for estimation of causal effects with propensity scores.  

This paper considers in detail the setting where inference for causal effects relies on specification of an outcome model that relates $\ps$ to $Y_i$, but many strategies for estimating causal effects with propensity scores do not specify such an outcome model.  These include various matching and stratified estimators, and also inverse probability of treatment weighting (IPTW) procedures that utilize weighted estimating equations to estimate, for example, parameters in a marginal structural model \citep{robins_marginal_2000, lunceford_stratification_2004}.  These methods are, in a sense, inherently non-Bayesian since they do not rely on a likelihood specification.  As such, the discussion of Section \ref{fundamental} does not directly apply. 

However, quasi-Bayesian procedures such as that outlined in Section \ref{cutbayes} hold promise for incorporating Bayesian methods into estimation of causal effects with these non-likelihood-based strategies because $\Delta$ above can represent any estimator.  Upon definition of $\Delta$ (e.g., based on matching or weighting), quasi-Bayesian inference could follow from the posterior-predictive distribution of that estimator obtained using expression (\ref{cutout}) above, without specifying $p(Y_i|X_i, \pscut, \theta_{Y|X,e^*})$.  Variations of this strategy were employed in \cite{hoshino_bayesian_2008} and \cite{kaplan_two-step_2012}, and this is an important area for continued research.

\bibliographystyle{biom}
\bibliography{bayesps_arxiv.bib}

\label{lastpage}
\end{document}